\documentclass[12pt]{article}
\usepackage{amsmath, amsthm, amssymb, amsfonts}  
\usepackage{graphicx} 
\usepackage{epsfig}
\newcommand{\sect}[1]{\section{#1}\setcounter{equation}{0}}

\newcommand{\be}{\begin{equation}}
\newcommand{\ee}{\end{equation}}
\newcommand{\nn}{\nonumber}
\newcommand{\s}[1]{{\bf#1}}                  

\def\half{\frac{1}{2}}
\def\thalf{\tfrac{1}{2}}

\def\tr{\hbox{Tr}}
 \def\eqalign#1{%
\null\,\vcenter{\openup\jot\m@th
  \ialign{\strut\hfil$\displaystyle{##}$&$\displaystyle{{}##}$\hfil
      \crcr#1\crcr}}\,}
\begin{document}
\bigskip
 \centerline{\bf Causal Spin Foams}
\bigskip
\centerline{ Giorgio Immirzi  }
 \medskip
\centerline{ Colle Ballone, Montopoli di Sabina (Italy) }

\centerline{\footnotesize e-mail: giorgio.immirzi\@@pg.infn.it }

\medskip
\abstract{\footnotesize I discuss how to impose causality in spin-foam models, separating forward and backward propagation,  turning a given triangulation to a 'causal set'. 
I show that the criteria proposed to identify the forward, causal sector of the theory are equivalent. Essential to the argument is the closure coondition for each 4-simplex.
}

 \sect{Introduction}
 The spin-foam models of quantum gravity\cite{BarrettCrane} are an attempt to realize the idea of quantizing Regge Calculus \cite{regge}, discretizing space-time in a simplicial complex, and calculating a `partition function'
 integrating over all possible configurations. However, a partition function is not really an approximation to a path integral: it does not distinguish the past from the future. This is just like
  the Wheeler De Witt equation, which unlike the Schr\"odinger equation knows no time, 
and is expected to give  a `forward' and a `backward'
 propagation\cite{colosi}\cite{roWdW}.
 But I would argue that the two should be separated,
and that the forward is the causal one. 
 
 A symptom that something is missing to the model
 comes from the asymptotic analysis of the model of 
 Engle-Livine-Pereira-Rovelli\cite{ELPR}  and Freidel-Krasnov\cite{FK} performed by
J.Barrett et al.\cite{barrettetal} and Mu Xin Han et al.\cite{muxin}. A saddle-point expansion
gives in the limit for the contribution of each triangle $\Delta_{ab}$, shared by tetrahedra $e_a$ and $e_b$ in a 4-simplex $v$,  a Regge-like expression:
\be 
 N _{ab+}e^{i{\mathcal A}_{ ab} \Theta_{ab} } + N _{ab-}e^{  -i{\mathcal A}_{ ab}\Theta_{ab} } 
 \label{reggelimit}\ee
${\mathcal A}_{ab}$ the area of the triangle, $\Theta_{ab}$ the (hyperbolic) dihedral  angle between $e_a$ and $e_b$. The second term in (\ref{reggelimit})  is there because in  the expansion one finds two saddle points, related by a parity  reflection; very much like what one finds in the simple model studied in \cite{colosi}; this is the `cosine problem'\cite{marko}.

 Some time ago  D.Oriti and E.Livine\cite{oritilivine}, in an analysis of the original model\cite{BarrettCrane}, emphasized the importance of  introducing  
 an element of causality in the model; and indeed the approach to quantum gravity based on dynamical triangulations  became `causal dynamical triangulation'\cite{ambjorn} when it was decided that causality was the missing ingredient. 

A possible way to impose causality on a spinfoam model has been suggested by M.Cortes 
and L.Smolin\cite{smolincortes}, based on the work of W.Wieland\cite{wieland}.
Simplifying,  for each tetrahedron  $e$ one has 
a closure constraint for the area tensors $S_f^{IJ}$ of its triangles, and for each 4-simplex
$v$ a closure constraint for the volume vectors $V^I_{e}$ of its 
tetrahedra\cite{caselle} 
\be
\sum_{f\in e}S^{IJ}_f=0;\quad\quad \sum_{e\in v}V^I_{ev}=0
\label{clo}\ee
i.e. the sum of the oriented areas and of the oriented volumes must be zero; I shall define these
quantities more precisely below. It is also commonly assumed that all tetrahedra are  space-like\footnote{in a previous paper\cite{me} I worried that  time-like tetrahedra might be needed to model specific spacetimes . Ignoring the background one may well  decide a-priori their absence,  crucial to the argument that follows.},
 i.e. all the $V_e^I$ are time-like, the actual volume being
${\mathcal V}_e=\frac{1}{6}\sqrt{-V^I_eV_{eI}}$. But if the 
$V^I_e$ are assumed time-like, they can only sum to zero for each $v$  if some $V_e^0$ are positive, some negative; some tetrahedra must be oriented forward, some backward, the closure condition becoming a sort of Kirchoff law for each 4-simplex. A given triangulation, if this orientation is dictated a priori,  becomes a sort of `causal set' in the sense of
R. Sorkin\cite{rafael}, or an `energetic causal set' \cite{smolincortes};
a simple detailed  example will be shown in the last section of this paper.

In this way the model becomes  `causal'; but  to give a single term in 
(\ref{reggelimit}) the configurations over which one integrates have to be limited to the forward oriented ones; proposals on how to identify them have been made and investigated in \cite{muxin} \cite{engle}\cite{riello}\footnote{ another point of view\cite{riello} is that the second term in (\ref{reggelimit}) accounts for the contribution of `anti-spacetimes' fluctuations, regions of negative lapse function; from the point of view of this paper and of \cite{muxin}, 4-simplices with negative ${\mathcal V}_4$, separated from the rest by degenerate 
4-simplices. I find difficult to reconcile this with the overall causal structure of the model.}; 
I will show that the different formulations are equivalent. 
In \cite{muxin}  it is also shown  that (under appropriate non-degeneracy hypotheses) the configurations that satisfy the saddle point conditions  also satisfy the 4-simplex closure.  So it is not just 4-simplex closure, but  the preordained orientation
of all tetrahedra  what  limits the configurations over which one integrates. 

I assume  that I am given a triangulation of 4-space to a simplicial complex  ${\mathcal K}$,
with dual skeleton  the 2-complex $\mathcal C$, made of 4-simplices/vertices $v$, 
tetrahedra/edges $e$, triangles/faces $f$.
The only boundaries  of $\mathcal K$  are an intial and a final  triangulation of $S^3$;  the simplest case is the pentachoron, which has 5 vertices, for which I give details in the last section.

A combinatorial notion of orientation for a 4-simplex is given by 
an ordering of its vertices $(abcde)$, or $(P_1,...,P_5)$, which induces an orientation of its 5 tetrahedra\\
$\{(abcd), (abec), (abde),(aced),(bcde)\}$, that in turn determines the orientation of the triangles 
e.g. $(abcd): (bcd), (cad), (abd), (bac)$. With these rules, each triangle 
within a 4-simplex is in two tetrahedra with opposite orientation. 
Even permutations of vertices do not change orientation, odd ones reverse it.
${\mathcal K}$ must be \underline{orientable}, meaning that an order of the vertices can be chosen for each 4-simplex such that each tetrahedron  belongs to two 4-simplices with opposite orientation. For example:
\be
 \begin{matrix}
(abcde'):&  a b c d&a b e' c& \underline{a b d e'}& a c e' d&b c d e'\\
(abdc'e'):& a b d c'&\underline{a b e' d}&a b c' e'& a d e' c'& b d c' e'
\end{matrix}
\nn\ee
Regge's original idea\cite{regge} was that each 4-simplex $v$ is a chunk of 4-space with a flat inside,
curvature residing  in the {\bf bones} (triangles/faces) $f$;  $e^I_\mu$ is a tetrad 1-form in a  coordinate patch covering $v$; Lorentz tranformations connect the frames in the tetrahedron $e=v\cap v'$,  overlap of 
two 4-simplices. The spacetime curvature shows up when going round a bone with successive transformations one does not come back to the original frame. 

The 4-volume form $e^0\wedge e^1\wedge e^2\wedge e^3$ characterizes the 
`geometric' orientation of the 4-simplex;  integrated, it  gives for each 4-simplex a positive 4-volume ${\mathcal V}_{4v}$, and  each tetrahedron $e\in v$ a 3-volume 4-vector $V^I_{ev}=\int\epsilon^I_{\ JKL}e^J\wedge e^K\wedge e^L$.  The 4 triangles that bound each tetrahedron have area tensors $S^{IJ}_{f}$, with  the crucial property that $\eta_{IJ}V^I_{e v}S^{JK}_{f}=0$; these 
area tensors are chosen as independent variables instead of the   tetrads. 
Within a 4-simplex $(1,...,5)$ tetrahedra can be labeled by the vertex they do not include,
triangles by the tetrahedra they border. One can easily derive that
\underline{classically} for each 4-simplex, area tensors, 3-volume vectors and 4 volume are related as:
\be
  \epsilon^{IJKL}V_{e K}V_{e' L}=2{\mathcal V}_4S^{IJ}_{ee'};\quad  \epsilon_{IJKL}V_2^IV_3^JV_4^KV_5^L={\mathcal V}_4^3
\label{fund}\ee 
Since we  assume
all tetrahedra to be space-like,  all the 4-vectors $V^I_{ev}$ are time-like, pointing in opposite time directions for a tetrahedron  shared by $v$ and $v'$; this  links  the `combinatorial' 
 and the `geometric' notion of orientation.

\sect{ }
  
 For a given triangulation the spinfoam amplitude $A(K)$  is  the
 integral over all  holonomies $g_{vv'}\in SL(2,{\mathbb C})$ of the
  product of  the face amplitudes $A_f$ associated to each face/triangle\cite{carlo}.
   Each holonomy is factorized as $g_{vv'}=g_{ve}g_{ev'},\ g_{ve}=g_{ev}^{-1}$ 
   if $e=v\cap v'$, and a `simplicity projector'\cite{riva} is  inserted for each $e$ in the chain.
      For example,   for a face of 3 steps:
 \[ A_f=\tr\, P_jg_{e_1v_1}g_{v_1e_2}P_j g_{e_2v_2}g_{v_2e_3}P_j g_{e_3v_3}g_{v_3e_1}
\] 
 the trace taken in a rep. $(j_f,\gamma j_f)$ of SL(2,${\mathbb C})$, and   
\be
P_j=\sum_m|(j,\gamma j)jm><(j,\gamma j)jm|:=\sum_m|jm)(jm| .
\label{proj}\ee
This form reflects the key points of the EPRL model: one only sums over the `$\gamma$-simple'
 representations of  the SL(2,$\mathbb C$),
 i.e.  those with indices $(j,\gamma j)$, $j$ a (half)-integer, 
and over the lowest SU(2) within their decomposition;  the areas of the triangles 
 ${\cal A}_f=\sqrt{\half S^{IJ}_fS_{f\, IJ}}$ are quantized and given by $\gamma j_j$.

 Explicitely,   the general expression for the spinfoam amplitude is:
\begin{align}
A(K)&=\int\prod dg_{ev}\prod_f A_f=\cr
\quad &=\int\prod dg_{ve}\prod_f 
\sum_{j_f} d_{j_f}\sum_{m\,m'..}(j_fm|g_{ve}^{-1}g_{ve'}|j_fm')...(j_fm''|g^{-1}_{v'e''}g_{v'e}|j_fm)
\label{A(K)}\end{align}
where  link/tetrahedron  $e$ enters $v$, $e'$ leaves it. 
In the appendix  I show how to express the projectors in terms of coherent states, and to rewrite 
this expression as:
\begin{align}
A(K)&=\cr
=\sum_{j_f}&d_{j_f}    \int\prod dg_{ve}\prod_{v\in f} d{\s n}_{ef}
\frac{< \s n_{ef}|g_{ve}^{\dagger -1}g_{v'e}^{-1}| \s n'_{ef}>^{2j_f}}
{<\s n'_{ef}| g^{\dagger -1}_{v'e}g^{-1}_{v'e}|\s n'_{ef}>^{j_f(i\gamma+1)+1}\,
    <\s n_{ef}| g^{\dagger -1}_{ve}g^{-1}_{ve}|\s n_{ef}>^{-j_f(i\gamma-1)+1}}\cr
=\sum_{j_f} &d_{j_f}   \int\prod dg_{ve}\prod_{v\in f} d{\s n}_{ef}e^{iS}
\label{A(K)bis}\end{align}
The last line above prepares the ground for the saddle point analysis of the large $j$ 
behaviour, which we expect to be dominated by the `critical configurations', where  
$Re S=\frac{\partial S}{\partial g_{ve}}=\frac{\partial S}{\partial \s n_{rf}}=0$.
We now have two sets of integrals, over the link group element $g_{ev}$ and over the $\s n_{ef}$, which can 
be interpreted as normal to the triangle. The two sets give independent descriptions of the geometry,
 which are linked for  critical configurations which extremise $S$. 
 
For a 4-simplex $v$ the volume vectors of the five tetrahedra can be taken, up to a 
proportionality constant, as
\be
V^I_{ve}=\epsilon_{ve}g^I_{ve J}{\mathcal T}^J
\label{V}\ee
 where  ${\mathcal T}^I=(1,0,0,0)$, and $\epsilon_{ve}=\pm 1$ determines the orientation of the 
 tetrahedron in $v$,  which  must be preassigned;
if $e=v\cap v',\ \epsilon_{ve}=-\epsilon_{v'e}$. By eq.(\ref{fund}) these 
vectors  determine the 4-volume of the 4-simplex ${\mathcal V}_4$, and   the area tensors of the triangles
$S^{IJ}_{(ee')}$. On the other hand from (\ref{A(K)bis}) we see that the
quantum theory gives the triangles quantized areas, with unit normals $\s n_{ef}$, so that 
\be
 { ^*\!}S^{IJ}_{f}={\mathcal A}_{f}\,g_{ ve K}^Ig_{ve L}^J
 \big({\mathcal T}\wedge (0,\s n_{ef})\big)^{KL}
\label{VNS}\ee
 Do these two descriptions agree?  Eq. (\ref{fund}) written for 
 ${ ^*\!}S^{IJ}_{ee'}$ reads
 \be
  V_{e}^IV_{e'}^J-V_{e}^J V_{e'}^I=-2  {\mathcal V}_4{ ^*\!}S^{IJ}_{ee'}
 \label{VVSdual}\ee
 If we multiply this equation by ${ ^*\!}S_{ee' IJ}$ and replace in it (\ref{VNS}) we obtain:
 \begin{align}
 {\mathcal V}_4 {\cal A}_{ee'} ^2&=
 {\cal A}_{ee'}\,
 \epsilon_{e}\epsilon_{e'}
 (g^I_{e v\,K} g^J_{e' v\,L}- g^J_{e v\,K} g^I_{e' v\,L} ) {\mathcal T}^K{\mathcal T}^L
 \,\eta_{IM}\eta_{JN}     g^M_{e v\,P}g^N_{ev\,Q}
 \big({\mathcal T}\wedge (0,\s n_{ee'})\big)^{PQ}=\cr
 &= {\cal A}_{ee'}\,
 \big(- \epsilon_{e}\epsilon_{e'}
 ( g^{-1}_{e v} g_{e' v})^\alpha_{\ 0} n^\alpha_{ee'}\big)
 \label{key}\end{align}
 which is my key equation.
 According to J.Engle, to get  the `proper vertex amplitude'\cite{engle}, with the desired asymptotic behaviour\cite{engle2},
 one should limit the integrations to  the configurations  for which the RHS of this equation is positive.
This agrees nicely with what MuXin Han et al. find\cite{muxin}: in the forward time-oriented sector of the theory the 4-volumes $ {\mathcal V}_4$ of all 4-simplices are positive. We see therefore that the criteria proposed to identify the forward, causal sector of the theory are equivalent.

In conclusion, to give a causal structure to a spin-foam theory we must use orientable triangulations,  with a-priori given orientations for the tetrahedra, and limit the integrations to give positive 4-volumes to all 4-simplices, or equivalently to  
 proper vertex amplitudes.

 \sect{ An example: the evolution of the pentachoron.}
 
 The `pentachoron' is a simple model for $S^3$, with an exotic name: five points all connected to each other.  One could think of more ambitious models along the same lines,  with 19 or 124 vertices\cite{kitchensink}\cite{ruth}. Following the rules explained in\cite{kitchensink},
 a triangulation evolving a pentachoron $(abcde)$ at t=0 
to a later pentachoron $(a'b'c'd'e')$ at t=1  can be realized connecting with edges all the 
vertices of the first to the vertices of the second, but \underline{omitting} the edges
$(aa'),(bb'),(cc'),(dd'),(ee')$; in this way we realize a division of the 
spacetime $S^3\otimes R$ between t=0 and t=1 in 30 4-simplices, . 
This triangulation can be proved to be orientable,
i.e. it can be organized  so that each of the 70 tetrahedra is in two 4-simplices with
opposite orientation, for ex. 
\be
 \begin{matrix}
(abcde'):&  a b c d&a b e' c& \underline{a b d e'}& a c e' d&b c d e'\\
(abdc'e'):& a b d c'&\underline{a b e' d}&a b c' e'& a d e' c'& b d c' e'
\end{matrix}
\label{tetras}\ee
and can therefore be described by a graph like the one below;
this graph is meant to explain the sense
in which this evolution can be regarded as a mini-causal-set: the pentagons represent 4-simplices, the oriented lines linking them the tetrahedra they share or, in the dual interpretation,  
the discrete spin connection $g_{vv'}$: 
 
  \includegraphics[width=\linewidth]{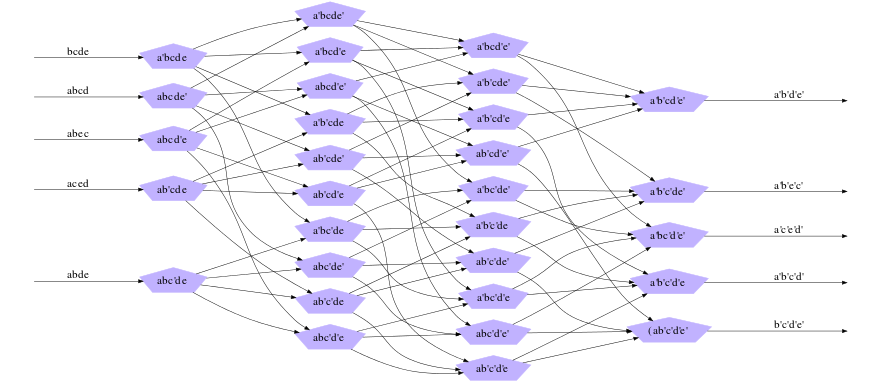}
  
However, one should not take it too literally;  the initial pentachoron (abcde) comes before 
the final (a'b'c'd'e'), but that notion does not apply to the intermediate stages; there is no sense in which (abcde') comes  `before'  (abdc'e'). This may well be the more interesting lesson to be drawn from  the example.

I would like to thank Carlo Rovelli for some crucial suggestions on an earlier version of the manuscript,
 Dimitri Marinelli for help with the graph,  Antonia Micol Frassino for suggesting improvements,  Sachindeo Vaidya  for the warm hospitality at the Indian Institute of Science, and Sumati Surya  for the hospitality at the Raman Research Institute 
in Bangalore.

\bigskip
\noindent{\bf Appendix: The spinfoam amplitude}

In this appendix I shall give a quick derivation of the expression for the spinfoam amplitude, following
\cite{carlo}\cite{muxin}.
One uses the unitary irreducible representations $(k,\nu)$ of SL(2,${\mathbb C}$), which act 
as $g\,\triangleright f(z)=f(z\,g)$
on functions of the spinor $z=\left(\begin{matrix}z_0\cr z_1\end{matrix}\right)$
such that $ f(\lambda z)=\lambda^{-1+i\nu+k}\bar\lambda^{-1+i\nu-k}f(z)$.
 The Hilbert space ${\mathcal H}_{k,\nu}$ of such functions  
can be realized as the space  of the functions  
$f\in L_2(SU(2)$ such that $ f(e^{i\varphi\sigma_3}u)=e^{2ik\varphi}f(u)$.  
A complete orthonormal set is: $\{|(k,\nu);jm>\},\  j\geq k,\ m\in(-j,j)$.

Writing         
 $g=\left(\begin{matrix}a&c\cr b&d\end{matrix}\right)=
kh=\left(\begin{matrix}\lambda^{-1}&\mu\cr
0&\lambda\end{matrix}\right)\left(\begin{matrix}\bar \beta&-\bar \alpha\cr \alpha&\beta\end{matrix}\right)$,  $h\in SU(2)$,  and defining: \\
$ug:=k(g,u)h(g,u)$, the action of $g\in SL(2,{\mathbb C})$ on the basis for this space will be: 
\be
<u|g|(k,\nu);jm>= \sqrt{d_j}\, \lambda(g,u)^{-k+i\nu-1}\bar\lambda(g,u)^{k+i\nu-1} D^j_{km}\big(h(g,u)\big)
 \label{gonD}\ee
with $d_j=2j+1$,
 having taken  $(\alpha,\beta)$ as the spinor, and 
 ${\mathcal H}_{k,\nu}=\bigoplus_{j\geq k}{\mathcal H}_j$, ${\mathcal H}_j$ the space on which the j-th representation of SU(2) acts.  In the expressions that follow the  $\lambda$-s  appear in pairs and may be taken real.
 The `simplicity constraints'  limit the representations that appear in the amplitude to be
 of the `$\gamma$-simple' type $(k,\nu)=(j,\gamma j)$, and the SU(2) representation to be the lowest. 
 With these preliminaries, in the expression (\ref{A(K)}) for the ELPR/FK spinfoam amplitude,
using the (\ref{gonD}) and inserting a complete set, each matrix element can be written as:
 \begin{align}
 (jm|g^{-1}g'|jm')&=d_j\int_{SU(2)} du\, \frac{\overline{ D}^j_{jm}(h(g,u))D^j_{jm'}(h(g',u))}
{\lambda(g,u)^{2i\gamma j+2}\lambda(g',u)^{-2i\gamma j+2}}=\cr
&=d_j\int_{SU(2)} du\frac{<jm|h(g,u)^\dagger|jj><jj|h(g',u)|jm'>}
{\lambda(g,u)^{2i\gamma j+2}\lambda(g',u)^{-2i\gamma j+2}}
\end{align}
therefore  in the expression of $A_f$ the trace of the product can be rearranged as a product of matrix elements
of the form:
\begin{align}<jj|h(g',u')h(g,u)^\dagger|jj>&=<\tfrac{1}{2}\tfrac{1}{2}|h(g',u')h(g,u)^\dagger|\tfrac{1}{2}\tfrac{1}{2}>^{2j}=\cr
&=<\tfrac{1}{2}\tfrac{1}{2}|k(g',u')^\dagger u'g'^{\dagger -1}
\,g^{-1}u^\dagger k(g,u)|\tfrac{1}{2}\tfrac{1}{2}>^{2j}=\cr
&=\left(\frac{<\tfrac{1}{2}\tfrac{1}{2}|u'g'^{\dagger -1}
\,g^{-1}u^\dagger |\tfrac{1}{2}\tfrac{1}{2}>}{\lambda(g',u') \lambda(g,u) }\right)^{2j}=
\left(\frac{<\s n'|g'^{\dagger -1}
\,g^{-1}|\s n>}{\lambda(g',u') \lambda(g,u) }\right)^{2j}.
\nn\end{align}
 Here I have  used: 
$k(g,u)|\thalf\thalf>=\lambda(g,u)^{-1}|\thalf\thalf>$,  
$\lambda(g,u)= [<\thalf\thalf|u g^{\dagger -1}g^{-1}u^\dagger|\thalf\thalf>]^{1/2} $,
and defined the coherent states $| \s n>:=u^\dagger|\thalf\thalf>$
with $\sigma\cdot\s n=u\sigma_3u^\dagger$. Replacing these expressions in (\ref{A(K)}),
 we
obtain the equation (\ref{A(K)bis}) given in the text:
\[A(K)=\sum_{j_f}d_{j_f}    \int\prod dg_{ve}\prod_{v\in f} d{\s n}_{ef}
\frac{< \s n_{ef}|g_{ve}^{\dagger -1 }g_{v'e}^{-1}| \s n'_{ef}>^{2j_f}}
{\lambda(g_{v'e},u')^{2j_f(i\gamma+1)+2}\,\lambda(g_{ve},u)^{-2j_f(i\gamma-1)+2}}
\]

\begingroup\raggedright 
\endgroup
 \end{document}